\pdfoutput=1
\documentclass{aastex701}
\submitjournal{ApJ}

\begin{document}

\title{Testing an Ap-like Magnetic Braking Origin for the Extended Main Sequence in Young Open Clusters}

\author[orcid=0000-0002-3084-5157,sname='Li']{Chengyuan Li}
\affiliation{School of Physics and Astronomy, Sun Yat-sen University, Daxue Road, Zhuhai, 519082, China}
\affiliation{CSST Science Center for the Guangdong-Hong Kong-Macau Greater Bay Area, Zhuhai, 519082, China}
\email[show]{lichengy5@mail.sysu.edu.cn}  

\begin{abstract}
Young open clusters commonly exhibit extended or split upper main sequences, which are widely interpreted as signatures of broad, and in some cases bimodal, stellar rotation distributions. The physical origin of the component commonly associated with slow rotation in this framework remains debated. One proposed channel invokes merger-induced magnetic braking analogous to that seen in Ap stars, which predicts that a large fraction of the relevant stars should host strong, stable surface magnetic fields and exhibit the Ap-like chemical peculiarity responsible for the characteristic $5200\,\mathrm{\AA}$ flux depression. We test this prediction using Gaia XP spectra of A-type stars in a sample of eight young open clusters and a diagnostic of the $5200\,\mathrm{\AA}$ flux depression. If Ap-like magnetic braking made a dominant contribution to the extended main-sequence phenomenon, a substantially enhanced incidence of $5200\,\mathrm{\AA}$ depression would be expected. Instead, we find no evidence for such a large excess. These results disfavour Ap-like magnetic braking as the dominant explanation for the extended main sequence in young open clusters, while leaving open magnetic or non-magnetic channels that would not produce a clear Ap-like $5200\,\mathrm{\AA}$ signature.
\end{abstract}



\section{Introduction} 
Young star clusters are no longer viewed as simple stellar populations (SSPs) in the strict sense. { Since the first report by \citet{Mackey2007},} high-precision photometry has shown that many young and { intermediate-age} clusters, in both the Magellanic Clouds and the Milky Way, exhibit extended main-sequence turnoffs and split upper main sequences \citep{Milone2009,Cordoni2018}. These structures have become one of the clearest manifestations of complexity in ostensibly SSPs and provide a sensitive probe of the physics that shapes the evolution of intermediate- and high-mass stars in cluster environments \citep[see review in ][]{Li2025}. Understanding the origin of the extended main sequence (eMS) is therefore important not only for interpreting cluster color--magnitude diagrams (CMDs), but also for constraining the roles of stellar rotation, binarity, magnetism, and related processes in young stellar populations.

Within this context, stellar rotation has become the most widely discussed framework for interpreting eMSs and split main sequences (MS) in young clusters. Early studies showed that rotational effects can naturally broaden or bifurcate the upper MS in young stellar populations, and subsequent work has further demonstrated that rotating stellar models can reproduce many of the observed photometric properties of these systems, including the morphology of the eMS and its dependence on cluster age \citep{BastiandeMink2009,Yang2013,BrandtHuang2015,Georgy2019}. This picture is consistent with { the standard rotational interpretation that the rMS is dominated by rapid rotators, while the blue main sequence (bMS) is composed of slowly rotating or nearly non-rotating stars} \citep{Dupree2017,Marino2018,Sun2019}. However, although rotation successfully explains much of the observed morphology, it does not by itself identify the physical origin of the population commonly associated with slower rotation in this framework. In particular, the mechanism capable of producing such a component in sufficient numbers, and on the relevant timescales in young cluster environments, remains uncertain.

Several scenarios have been proposed to explain the population commonly associated with slower rotation in the rotational framework, including tidally interacting binaries \citep{D'Antona2015,D'Antona2017} and disc-locking during the pre-main-sequence phase \citep{Bastian2020}. { These alternatives have already been examined in several recent studies. Using numerical simulations, \citet{Wang2023} showed that the standard tidal synchronization timescale is generally too long for most binaries to become efficiently locked in young clusters, and that the systems that can be synchronized on the relevant timescale are predominantly high-mass-ratio binaries. This is difficult to reconcile with the observational result that the slowly rotating population lies close to the single-star or low-mass-ratio binary sequence. Consistently, \citet{He2023} carried out time-domain spectroscopic observations of NGC~2422 and found that the fraction of stars showing significant radial-velocity variations is too small to support a slow-rotator population dominated by tidally braked binaries. By contrast, \citet{Bu2025} reported that disc-less protostars in the star forming region, NGC 2264, rotate systematically faster than disc-bearing ones, providing empirical support for the idea that star--disc coupling can play an important role in early spin-down.}

{ In contrast to these scenarios, the merger-induced magnetic-braking picture proposed by \citet{Wang2022}, building on the merger framework explored by \citet{Schneider2019}, has not yet, to our knowledge, been subjected to a direct observational test. This makes it particularly attractive in the present context, because it links the putative slow-rotator population to a specific and testable physical channel.} Magnetohydrodynamic simulations by \citet{Schneider2019} showed that mergers of massive stars can produce blue stragglers with surface magnetic fields of several kG, comparable to those observed in strongly magnetic early-type stars \citep{Shultz2019,Sikora2019}. \citet{Wang2022} suggested that this merger-induced strong-magnetism channel, originally studied for massive O-type stars, may also operate in MS A-type stars and could account for the bMS population in young clusters. { Here and throughout this paper, however, ``magnetic braking'' does not refer to the classical spin-down mechanism operating in cool stars below the Kraft break, but to a distinct process relevant to radiative-envelope A-type stars above it. The stars considered here are therefore not expected to undergo the familiar solar-type magnetic braking driven by a convective-envelope dynamo. Instead, the relevant mechanism requires a pre-existing strong, globally organized magnetic field of the kind observed in classical Ap stars, i.e., chemically peculiar A-type stars with stable large-scale magnetism \citep{Braithwaite2004,Donati2009}. Such fields are commonly interpreted as fossil in origin, although merger-related formation channels have also been proposed \citep{Ferrario2009,Wickramasinghe2014}. In this picture, binary mergers produce stars with strong, stable magnetic fields that are subsequently spun down by magnetic braking, thereby linking binary interaction, magnetism, and rotational evolution within a single evolutionary pathway. }


A key prediction of this scenario is that, if merger-induced magnetism and subsequent magnetic braking contribute substantially to the observed eMS morphology, then the relevant stellar population should show an enhanced incidence of Ap-like chemical peculiarity. { This expectation arises because the strong, stable, globally organized magnetic fields invoked in this scenario are also the defining property of classical Ap stars, whose stable radiative atmospheres commonly develop characteristic diffusion-driven chemical peculiarities.} In particular, the flux depression around \(5200\,\mathrm{\AA}\) is a classical signature of chemically peculiar Ap stars and provides a well-established observational diagnostic of this { phenomenon \citep{Maitzen1976,Paunzen2005,Paunzen2022,Xie2025}}. The \cite{Wang2022} scenario therefore predicts not merely an enhanced incidence of Ap-like signatures among bMS stars relative to a normal population of young A-type stars, but a high absolute fraction if this channel is responsible for the bMS itself. In that case, stars associated with the bMS should be dominated by objects showing Ap-like characteristics, so that the expected incidence of such signatures would be of the same order as the bMS (or slow-rotator) population fraction rather than only marginally elevated above the normal baseline. This makes the \(5200\,\mathrm{\AA}\) depression a direct empirical test of whether Ap-like magnetic braking plays a major role in producing the eMS in young clusters.

Testing this prediction directly through magnetic-field measurements is, however, challenging \citep{Donati2009,Wade2016}. Large-scale spectropolarimetric observations of A-type stars in multiple young clusters remain observationally expensive and are not yet available for the kind of homogeneous population-level analysis required here. Gaia XP spectra now provide a powerful alternative by delivering uniform low-resolution spectrophotometry for large samples of cluster stars across the Galaxy, including the wavelength region around \(5200\,\mathrm{\AA}\) \citep{GaiaCollaboration2023,DeAngeli2023,Montegriffo2023}. This makes it possible to search, in a consistent statistical manner, for Ap-like signatures among A-type stars in multiple young Galactic open clusters, and thereby to test whether such stars are sufficiently common to support an Ap-like magnetic-braking origin for the eMS.

In this paper, we present a population-level test of this scenario using Gaia XP spectra for stars in eight young Galactic open clusters. An empirical flux-ratio diagnostic sensitive to the \(5200\,\mathrm{\AA}\) flux depression is constructed from reference samples of Ap stars and normal A stars, and a mixture model is then used to infer the fraction of Ap-like stars among the cluster A-type populations. The inferred Ap-like fractions are generally low, indicating that stars with Ap-like spectral signatures are unlikely to be sufficiently common to account for the eMS phenomenon in bulk. The results therefore do not support Ap-like magnetic braking as the dominant channel for producing the eMS while still allowing for the possibility that magnetic stars contribute in individual cases or in a minority subpopulation. The remainder of this paper is organized as follows. Section~2 describes the data and methodology, including the cluster sample, the Gaia XP spectra, the reference samples, the \(5200\,\mathrm{\AA}\)-based diagnostic, and the mixture-model framework. Section~3 presents the inferred Ap-like fractions for individual clusters and for the full sample. Section~4 discusses the implications of these results for the origin of the eMS, together with relevant caveats and limitations. Section~5 summarizes the main conclusions.



\section{Data and Method} \label{sec2}
\subsection{Cluster sample}
We selected eight nearby young open clusters from the membership catalog of \cite{Hunt2023}. Cluster members were taken directly from their publicly available Gaia-based kinematic membership catalog. Rather than applying a strict algorithmic selection, we chose clusters empirically to satisfy several practical conditions: a sufficiently rich stellar population, relatively low extinction, small enough distance that Gaia XP spectra cover most stars on the upper main sequence, and a clearly eMS whose upper part is dominated primarily by B- and A-type stars. The resulting sample spans \(\log(\mathrm{age/yr}) \approx 7.75\)--8.38 (\(\sim 60\)--240 Myr). The basic properties of the cluster sample are summarized in Table~\ref{tab:cluster_sample}.

\begin{deluxetable}{lccccc}
\tablecaption{Basic Properties of the Cluster Sample\label{tab:cluster_sample}}
\tablehead{
\colhead{Cluster} & \colhead{MOD50} & \colhead{logAge50} & \colhead{AV50} & \colhead{$N_{\rm eMS}$} & \colhead{$N_{\rm mem}$}
}
\startdata
M45         & 5.590 & 8.085 & 0.102 &  43 & 1721 \\
Melotte 20  & 6.185 & 7.745 & 0.242 &  49 &  938 \\
NGC 1039    & 8.390 & 8.087 & 0.143 &  44 & 1095 \\
NGC 2287    & 9.210 & 8.189 & 0.086 &  94 &  948 \\
NGC 2516    & 7.974 & 8.098 & 0.205 & 105 & 3784 \\
NGC 3114    & 9.916 & 8.220 & 0.137 & { 222} & 1624 \\
NGC 3532    & 8.280 & 8.376 & 0.120 & 220 & 3414 \\
NGC 6475    & 7.148 & 8.232 & 0.208 &  79 & 1427 \\
\enddata
\tablecomments{Columns MOD50, logAge50, and AV50 list the 50th-percentile catalog values reported by \cite{Hunt2023}. $N_{\rm mem}$ is the total number of members in the \cite{Hunt2023} membership catalog, and $N_{\rm eMS}$ is the number of stars in our adopted eMS sample.}
\end{deluxetable}

\begin{figure*}[t]
\centering
\includegraphics[width=\textwidth]{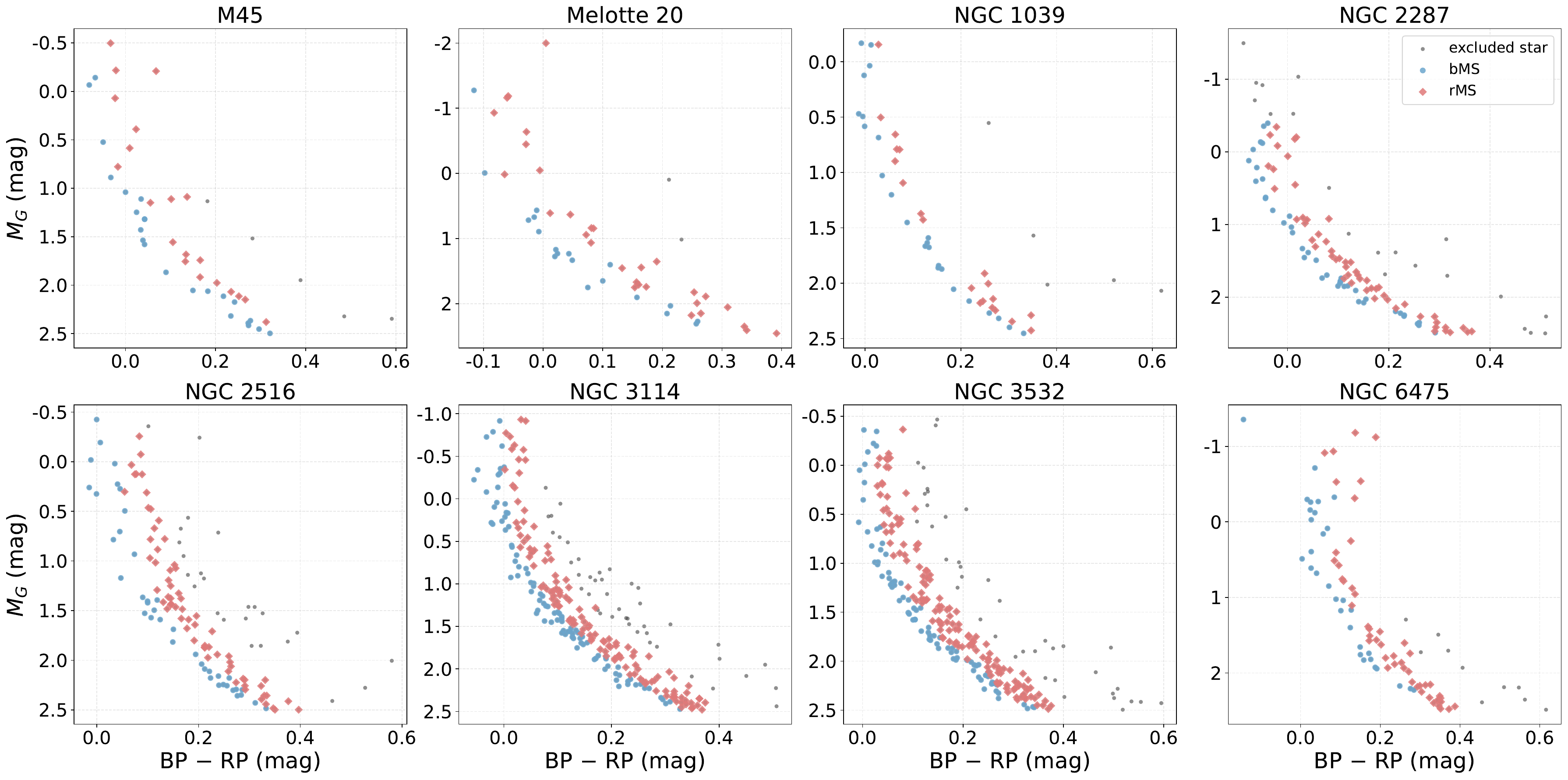}
\caption{Dereddened CMDs of the eight open clusters analyzed in this work. The sample is drawn from the astrometrically selected cluster catalog of \cite{Hunt2023}. We retain nearby, low-extinction systems with ages of $\sim 60$--240 Myr, sufficiently rich upper-MS populations, and clearly broadened or split MS morphologies. The eMS stars used in the analysis are defined empirically from the observed MS morphology, excluding the redder sequence dominated by likely binaries{ ; gray points mark stars excluded from the final eMS sample, including the redder sequence and a small number of obvious red outliers, while blue and red symbols indicate the adopted bMS and rMS subsamples, respectively, as shown in the legend}.}
\label{fig:cluster_cmd}
\end{figure*}

Figure~\ref{fig:cluster_cmd} shows the dereddened CMDs of the eight clusters and illustrates how we define the adopted eMS sample from the observed upper-MS morphology. We focus on stars in the eMS region, corresponding predominantly to A-type stars over the adopted magnitude range, while avoiding the redder loci dominated by likely high-mass-ratio binaries. Only in the youngest cluster does the selection extend up to \(M_G \sim -2\); in the remaining clusters, the bright limit is typically around \(M_G \sim -0.5\), so the sample includes at most a small number of late-B stars. For each cluster, we divide the CMD into magnitude bins and define the blue-edge boundary from the bluest 5\% of stars in each bin. We then measure the color offset of each star from this blue-edge locus{ . Because the upper MS is intrinsically broadened by rotation, there is no single photometric criterion that can cleanly separate binaries from single stars over the full magnitude range, especially for systems with relatively small mass ratios. We therefore adopt an empirical but uniform procedure to exclude the reddest part of the distribution in each magnitude bin. For the later-type part of the sample, corresponding roughly to late-A/early-F stars, we exclude the reddest 15\% of stars in each bin; these objects are expected to be dominated by likely high-mass-ratio binaries. Toward brighter, earlier-type stars, this excluded fraction is gradually reduced to 5\% and in some cases to zero, because rapid rotation together with evolution toward the terminal-age main sequence increasingly blurs the photometric distinction between binaries and single stars. We also exclude a small number of obviously red outliers by visual inspection, so that the retained eMS locus follows the characteristic broadened morphology reported in previous studies.} The retained stars are further divided into blue-MS and red-MS subsamples according to their dereddened color offset from the blue-edge boundary.


\subsection{Gaia XP Spectra}

Gaia XP spectrophotometric data were retrieved for the cluster members from Gaia DR3 by matching their Gaia DR3 \texttt{source\_id}. Depending on the available Gaia DR3 XP data product, the final spectra used in this work were obtained in two ways. For part of the sample, sampled wavelength--flux spectra were directly retrieved either from the VizieR catalog \texttt{I/355/xpsample} or through the Gaia Archive Datalink product \texttt{XP\_SAMPLED}. For the remaining stars, spectra were reconstructed from Gaia DR3 XP coefficients in \texttt{gaiadr3.xp\_continuous\_mean\_spectrum} using the \texttt{calibrate} routine in GaiaXPy{ \footnote{\url{https://www.cosmos.esa.int/web/gaia/gaiaxpy}}}.

We applied only basic integrity checks to the retrieved spectra, including verification of monotonic wavelength ordering, finite and non-negative flux values, and adequate wavelength coverage at both the blue and red ends; files failing these checks were discarded and re-retrieved. No additional signal-to-noise cut or XP quality-flag selection was imposed at this stage, as the stars in all eight clusters are relatively bright; even in the most distant cluster, the analyzed stars have apparent magnitudes of approximately $G<12$ mag, for which the Gaia XP spectra are expected to have adequate signal-to-noise for our flux-ratio measurements. The downloaded spectra were otherwise used in their original form, without dereddening, normalization, smoothing, or resampling. Extinction corrections were applied only later in the analysis when flux ratios were measured. In total, 844 of the 872 adopted cluster stars yielded valid XP-based flux-ratio measurements and were retained for the subsequent analysis.

\subsection{Reference Samples}

To place the cluster eMS stars in context, we constructed two comparison samples: a magnetic Ap-star sample and a chemically normal A-star sample. In both cases, the reference stars were further restricted to occupy approximately the same region of the CMD as the cluster eMS population (Figure~2, left), so that the comparison was carried out over a similar part of parameter space.

\begin{figure*}[t]
\centering
\gridline{
\fig{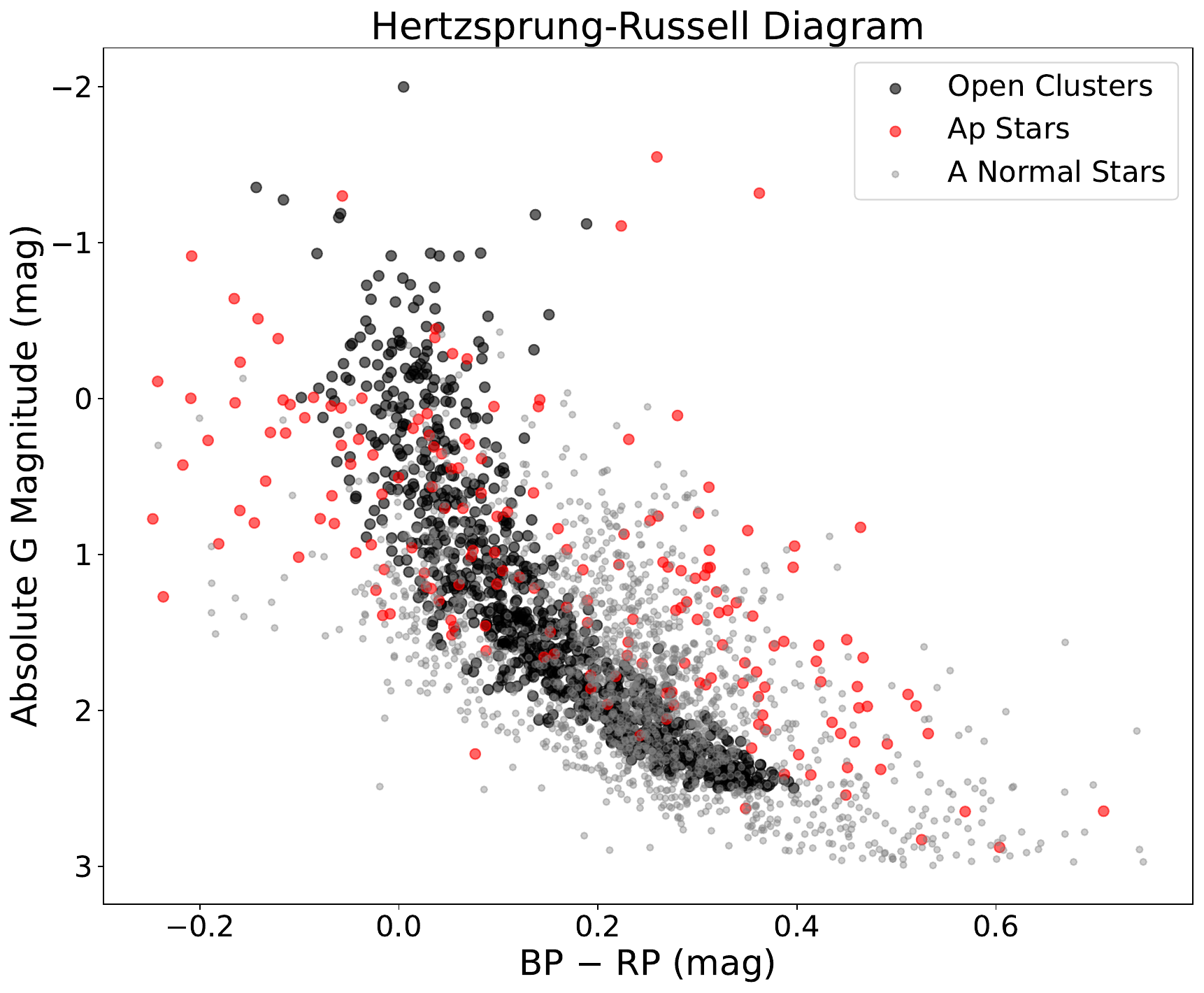}{0.44\textwidth}{}
\fig{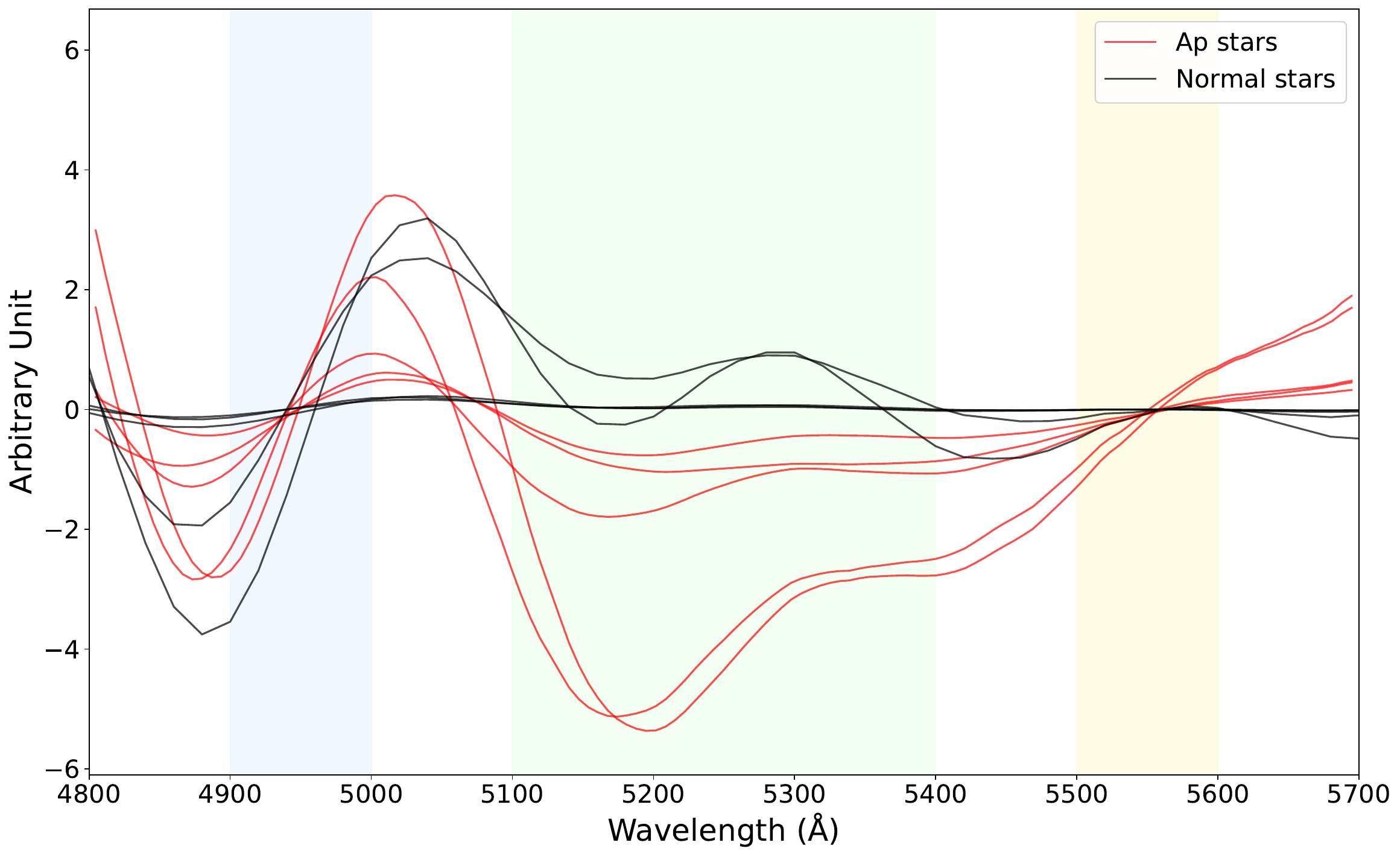}{0.56\textwidth}{}
}
\caption{Reference samples and definition of the empirical 5200\,\AA\ diagnostic used in this work. \emph{Left:} CMD of the empirical reference samples, including Galactic Ap stars from \cite{Scholz2019} and chemically normal A-type stars selected from GALAH DR3, as well as stars in the eight Galactic clusters. The reference stars are further restricted to the same CMD region occupied by the cluster extended MS population, so that the comparison is carried out over the same part of parameter space. \emph{Right:} Illustration of the Gaia XP flux-ratio diagnostic around the 5200\,\AA\ flux depression. The diagnostic is defined as the mean flux in the 5100--5400\,\AA\ band divided by the average of the mean fluxes in the 4900--5000\,\AA\ and 5500--5600\,\AA\ side bands, and is used to distinguish Ap-like stars from normal A-type stars in the cluster sample.\footnote{For visual comparison, the spectra shown in the right panel are locally linear-transformed so that the pseudo-continuum defined by the side bands is horizontal; the plotted vertical scale is therefore in arbitrary units. This visualization step is for display only and is not part of the flux-ratio definition used in the analysis.}}
\label{fig:reference_diagnostic}
\end{figure*}

\paragraph{Ap stars.}
The Ap-star reference sample was drawn from the strongly magnetic Ap-star catalog of \citet{Scholz2019}, which is based on Gaia DR2 data. The parent catalog contains 237 objects in total. From this catalog, we further restricted the sample to stars occupying approximately the same region of the CMD as the cluster eMS sample, as illustrated in Figure~\ref{fig:reference_diagnostic} (left), in order to ensure a more appropriate comparison. This yielded a working sample of 184 Ap stars. Among these, 150 were ultimately retained for the flux-ratio analysis, while the remaining 34 could not be matched to usable archive data.

\paragraph{Normal A stars.}
The chemically normal A-star reference sample was constructed from the { GALAH DR3 main star catalog \citep{Buder2021}}, { accessed via the Data Central TAP service} (\url{https://datacentral.org.au/vo/tap}). { We first selected stars with effective temperatures \(7500 \leq T_{\rm eff} \leq 10000\,\mathrm{K}\), surface gravities \(\log g > 3.5\), and metallicities \(-0.5 \leq [\mathrm{Fe}/\mathrm{H}] \leq 0.3\), in order to isolate mainly MS A-type stars with approximately near-solar metallicities. We further required \texttt{flag\_sp} \(\leq 1\) to retain stars with acceptable GALAH spectroscopic-parameter quality flags, and \(V_{\rm broad}>0\) to keep only stars with valid line-broadening measurements.} { For the hotter part of the sample, we applied additional abundance filters to remove stars with obvious chemical peculiarities. In particular, we required \([\mathrm{Si}/\mathrm{Fe}] \leq 0.3\), \([\mathrm{Cr}/\mathrm{Fe}] \leq 0.3\), and \([\mathrm{Sr}/\mathrm{Fe}] \leq 0.5\), motivated by the fact that these elements commonly show strong anomalies in magnetic chemically peculiar A/B stars \citep[e.g.][]{Ghazaryan2018}. We also removed likely \(\lambda\) Bootis candidates, identified empirically as stars satisfying both \([\mathrm{Mg}/\mathrm{Fe}] < -0.4\) and \([\mathrm{Al}/\mathrm{Fe}] < -0.4\), following the characteristic refractory-element deficiencies of \(\lambda\) Bootis stars \citep{Heiter2002}. These limits were not intended as formal classification boundaries, but were chosen empirically to balance the removal of chemically peculiar stars against the retention of chemically normal stars. We verified that changing these thresholds by \(\pm 0.1\) dex alters the final sample size by only about \(-5\%\) to \(+3\%\), indicating that our adopted cuts are reasonably stable.}

The resulting GALAH sample was then cross-matched with Gaia DR3 within \(2\arcsec\) to obtain Gaia \texttt{source\_id} values, producing a parent sample of 2187 stars. XP spectra were reconstructed for the matched stars, yielding 1962 objects with usable XP data. After applying extinction corrections, imposing the HRD cuts \((BP-RP)_0 \in [-0.25,\,0.75]\) and \(M_G \in [-2,\,3]\), { and performing a final empirical cleaning in the extinction-corrected CMD, we retained only stars lying within a broad envelope defined from the combined eMS populations of all eight clusters: specifically, stars were kept if their \(M_G\) values were within 0.5 mag of the full eMS magnitude range and their \((BP-RP)_0\) values were within 0.3 mag of the full eMS color range. This step was intended to remove obvious outliers while retaining stars occupying approximately the same color--magnitude region as the cluster eMS population shown in Figure~\ref{fig:reference_diagnostic} (left).} The final normal A-star comparison sample contained 1637 stars.

\subsection{The 5200\,\AA\ Diagnostic}

The broad flux depression around \(5200\,\AA\) is a classical empirical signature of chemically peculiar Ap stars and is commonly interpreted as the result of enhanced line blanketing in atmospheres with Ap-like chemical anomalies, often associated with strong, stable magnetic fields \citep{Maitzen1976,Kupka2003,Shultz2019}. Because this feature is broad, it remains accessible at the low spectral resolution of Gaia XP spectra and can therefore be used as a practical empirical diagnostic for our cluster sample.

To quantify this feature in the XP data, we define an empirical flux-ratio index,
\[
R_{5200} \equiv \frac{F_{\rm center}}{(F_{\rm blue}+F_{\rm red})/2},
\]
where \(F_{\rm blue}\), \(F_{\rm center}\), and \(F_{\rm red}\) are the arithmetic mean fluxes measured over 4900--5000\,\AA, 5100--5400\,\AA, and 5500--5600\,\AA, respectively. In this work, the XP spectra are first corrected for extinction, and the index is then measured from the extinction-corrected spectra. By comparing the mean flux in the central band with the average level of the two side bands, this index captures the local depression around \(5200\,\AA\) while reducing sensitivity to the overall spectral slope. In this sense, \(R_{5200}\) may be regarded as a flux ratio, or equivalently as a pseudo-index that measures the relative strength of the local \(5200\,\AA\) depression.

This definition is designed for the characteristics of Gaia XP spectra. At XP resolution, the \(5200\,\AA\) depression is better treated as a broad spectral-shape diagnostic than as a narrow isolated feature, so a three-band flux ratio provides a simple and robust empirical measure. As shown by the Ap-star and chemically normal A-star reference samples (Figure~2, right), the two populations occupy systematically different, although partially overlapping, ranges of \(R_{5200}\). We therefore use this diagnostic in a statistical rather than star-by-star sense: the goal is not to assign a precise chemical classification to each cluster member, but to test whether the cluster subsamples show evidence for an Ap-like component and, if so, to infer its fractional contribution.

The uncertainty in \(R_{5200}\) is influenced by several factors. Random measurement scatter arises from the finite precision of the XP fluxes and from the limited spectral resolution of the reconstructed spectra. In addition, residual calibration errors, imperfect extinction correction, or small local continuum-shape mismatches may introduce systematic offsets in the measured ratio. The intrinsic strength of the \(5200\,\AA\) depression may also depend not only on chemical peculiarity, but to some extent on stellar parameters such as effective temperature. For this reason, the comparison is restricted to reference stars occupying approximately the same color--magnitude region as the cluster eMS population, so that the diagnostic is interpreted in a relative empirical sense within a closely matched parameter space. Finally, the locally linear transformation applied to the example spectra in Figure~\ref{fig:reference_diagnostic} (right) is for visualization only and is not part of the definition or measurement of \(R_{5200}\).


\subsection{Mixture Model}
\label{subsec:mixture_model}

To quantify the contribution of Ap-like stars to each cluster subsample, we model the observed distribution of \(R_{5200}\) as a mixture of the Ap-star and normal-A-star reference distributions introduced above. For each star, we first evaluate the relative support for the two reference populations through the Bayes factor
\begin{equation}
BF \equiv \frac{p(R_{5200}\mid \mathrm{Ap})}{p(R_{5200}\mid A_{\rm normal})},
\end{equation}
where \(p(R_{5200}\mid \mathrm{Ap})\) and \(p(R_{5200}\mid A_{\rm normal})\) denote the probability density functions of the Ap and normal-A reference samples, respectively. By construction, \(BF>1\) indicates that the measured \(R_{5200}\) is more consistent with the Ap reference distribution, while \(BF<1\) favors the normal-A reference distribution. We stress, however, that this statistic is used only to characterize the ensemble behavior of a subsample, rather than to assign an Ap/non-Ap class to individual stars.

The reference density functions are constructed non-parametrically using kernel density estimation (KDE) with \texttt{scipy.stats.gaussian\_kde}, based on the empirical \(R_{5200}\) distributions of the Ap and normal-A control samples. { The KDE bandwidth was selected by leave-one-out cross-validation over trial values between 0.005 and 0.05. This range was chosen empirically to span plausible smoothing scales for the observed \(R_{5200}\) distributions.} Once determined, these reference distributions were held fixed in all subsequent mixture-model fits.

For a given subsample, we assume that the observed \(R_{5200}\) distribution follows
\begin{equation}
p(R_{5200}) = \pi\, p(R_{5200}\mid \mathrm{Ap}) + (1-\pi)\, p(R_{5200}\mid A_{\rm normal}),
\end{equation}
where \(\pi\) is the fraction of stars with Ap-like 5200\,\AA\ behavior. We fit this model separately to the combined cluster sample and to the color-selected subsamples defined by \texttt{color\_diff}: bMS (\(-0.035 \le \texttt{color\_diff} < 0.030\)) and rMS (\(0.030 \le \texttt{color\_diff} < 0.267\)). { These bMS and rMS subsamples correspond to the blue- and red-main-sequence populations highlighted in Figure~\ref{fig:cluster_cmd}. The division at \texttt{color\_diff} \(=0.03\) was adopted empirically; with this definition, the relative numbers of bMS and rMS stars in all eight clusters remain within \(\sim 2/3\) to \(1\), broadly consistent with those found in our previous studies \citep{Li2017,Sun2019}. The adopted intervals span the full observed \texttt{color\_diff} range of the eMS population in our sample, so that the eMS stars are partitioned into blue and red components without introducing an additional intermediate subsample.}

Our primary estimate of \(\pi\) is obtained by maximizing the likelihood on a grid of 200 points over \(0.001 \le \pi \le 0.5\), and we adopt the maximum-likelihood value as the best-fit Ap-like fraction. { The lower bound was used as a numerical approximation to \(\pi=0\), while the upper bound was adopted as a conservative physical limit: in the framework of \cite{Wang2022}, even the extreme case in which all bMS stars are Ap-like would imply an Ap-like fraction of about 50\% in the full sample.}

\section{Results} \label{subsec:S3}

\subsection{$R_{5200}$ distributions of the color-defined eMS subsamples}

\begin{figure*}
\centering
\includegraphics[width=\textwidth]{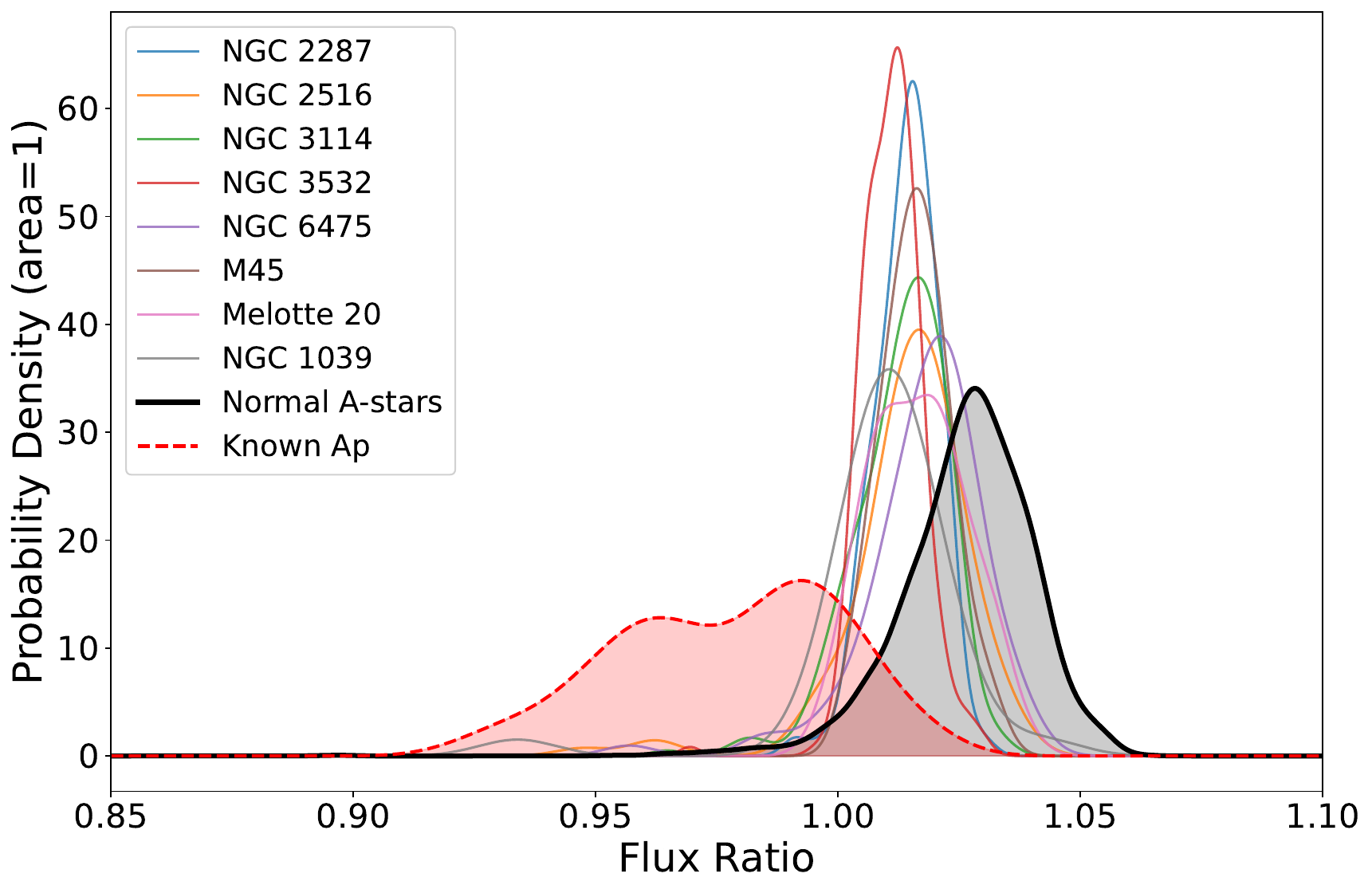}
\caption{
Distributions of the 5200\,\AA\ diagnostic $R_{5200}$ (Flux Ratio) for the color-defined eMS subsamples in the eight clusters. In each panel, the observed cluster distribution is compared with the empirical reference distributions of Ap stars and chemically normal A-type stars defined from the calibration samples. The cluster subsamples generally lie closer to the normal-A reference distribution than to the Ap reference distribution, although some show modest extension toward the Ap-like regime, more often in the bMS than in the rMS. This qualitative comparison motivates the mixture-model analysis presented below.
}
\label{fig:r5200_dist}
\end{figure*}

Figure~\ref{fig:r5200_dist} compares the observed $R_{5200}$ distributions of the color-defined eMS subsamples in the eight clusters with the empirical reference distributions defined by the Ap and normal-A calibration samples. As established in Section~2, the two reference populations occupy overlapping but systematically different regions in $R_{5200}$ space, providing an empirical basis for interpreting the cluster measurements in a statistical sense. Across the full cluster sample, the eMS subsamples are generally distributed closer to the normal-A reference distribution than to the Ap reference distribution, although some extension toward the Ap-like regime is present in several cases.

This qualitative behavior already suggests that a pronounced 5200\,\AA\ flux depression is not a dominant property of the eMS population as a whole. At the same time, the comparison does not indicate that the cluster subsamples are described by a purely normal-A population either, because in some clusters the distributions show modest tails or shifts toward the Ap reference locus. In particular, such extensions appear more often in the bMS subsamples than in the corresponding rMS subsamples, although the difference is not uniform from cluster to cluster and substantial overlap remains. We therefore treat Figure~\ref{fig:r5200_dist} as an empirical motivation for a quantitative population-level decomposition, rather than as evidence for star-by-star classification. 
In the next subsection, we formalize this comparison using the mixture model introduced in Section~2.5 and infer the Ap-like fraction, $\pi$, for the combined subsamples across all clusters.

\subsection{Inferred Ap-like fractions from mixture modeling}

Before turning to the population-level mixture inference, we first examine the star-by-star Bayes factor, which quantifies whether an individual star is more likely under the empirical Ap reference distribution or the normal-A reference distribution. In the combined cluster sample, only a minority of stars favor the Ap reference: { 22.6}\% have $BF>1$, while only { 7.0}\% and 1.9\% reach $BF>3$ and $BF>10$, respectively, with a median { $\log BF=-0.49$}. The bMS is somewhat more weighted toward the Ap reference than the rMS, with { 26.4\%, 10.9\%}, and 2.4\% of stars exceeding $BF>1$, $3$, and $10$, respectively, compared to { 20.1\%, 4.4\%}, and 1.6\% in the rMS; the median remains { $\log BF=-0.49$} for all three cases. Thus, although some stars show mild Ap-like tendencies, the star-by-star evidence remains weak overall and does not support a picture in which most eMS stars are individually identifiable as Ap-like.

\begin{figure*}
\centering
\includegraphics[width=\textwidth]{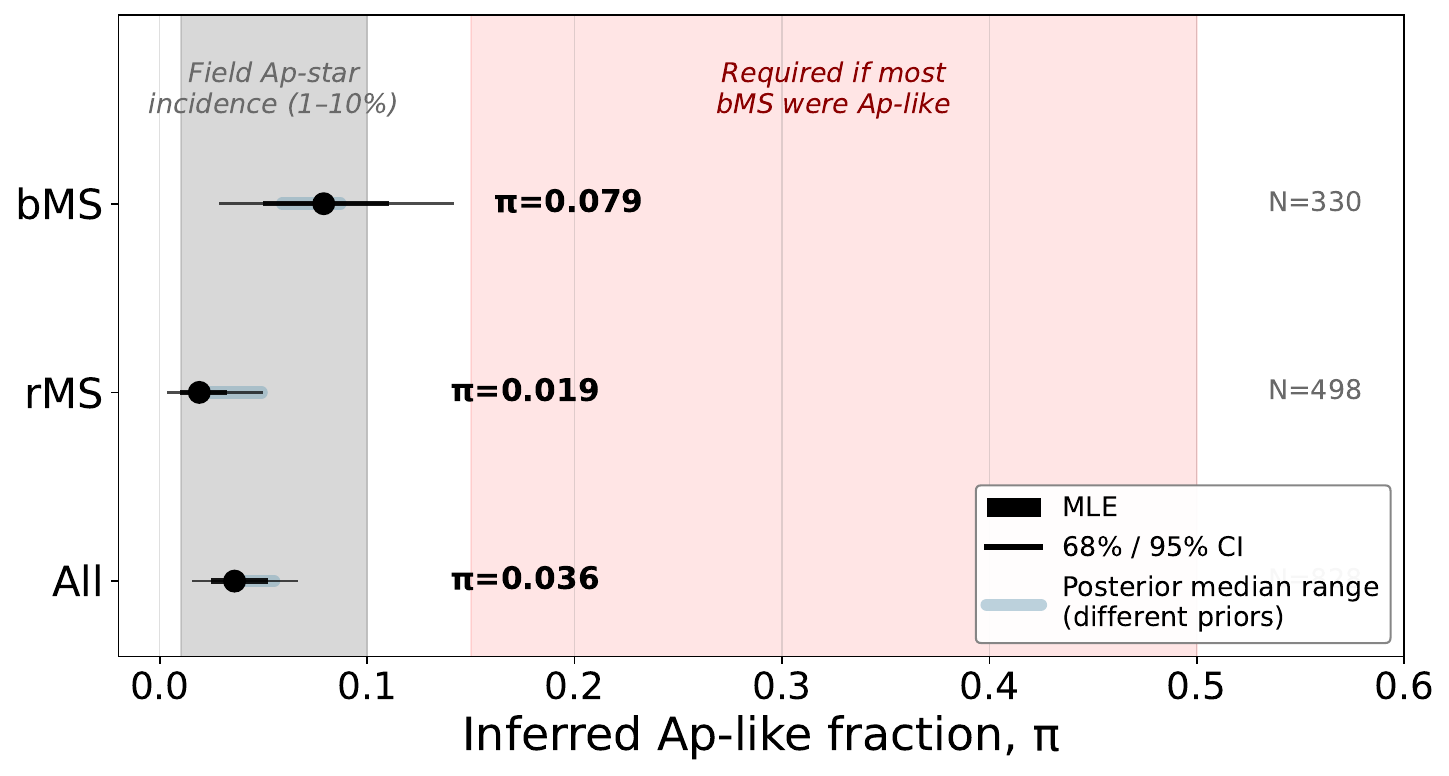}
\caption{Summary of the inferred Ap-like fraction for the full sample and for the bMS and rMS subsamples. Black points mark the maximum-likelihood estimates, while the thick and thin horizontal bars show the 68\% and 95\% confidence intervals, respectively. The gray shaded band indicates the reported incidence range of field Ap stars, and the red shaded band marks the approximate fraction required in the full sample if most bMS stars were Ap-like. Even for the bMS subsample, the inferred fraction remains well below that required level.}
\label{fig:ap_fraction}
\end{figure*}

This is precisely why we do not interpret $R_{5200}$ or $BF$ as a star-by-star classification tool. Instead, we use the mixture model introduced in Section~2.5 to ask whether the observed $R_{5200}$ distribution of a cluster subsample is better described as a population-level mixture of the Ap and normal-A reference distributions, and to infer the corresponding Ap-like fraction, $\pi$. The maximum-likelihood results for the combined cluster sample are summarized in Figure~\ref{fig:ap_fraction}. For all clusters combined, we infer a low Ap-like fraction, with a best-fit value of { $\pi=0.036$}. When the sample is separated by sequence, the bMS yields the larger value, { $\pi=0.079$}, while the rMS gives a substantially smaller value, { $\pi=0.019$}. Thus, the mixture-model inference is consistent with a modest enhancement of Ap-like stars in the bMS relative to the rMS, but it does not support a scenario in which Ap-like stars dominate either sequence.

The inferred fractions remain below the level that would be required if Ap-like stars accounted for a substantial fraction of the bMS population, and far below the level required if they dominated it \citep[e.g.,][]{Milone2016,Milone2017,Li2017,Sun2019}. At the same time, the inferred values are not grossly inconsistent with the incidence of Ap stars in the field, especially for the full and rMS samples, indicating that the observed $R_{5200}$ distributions do not require a large excess of chemically peculiar magnetic stars in the cluster population.

We emphasize that this conclusion is based primarily on the maximum-likelihood inference, with the Bayesian calculations used as a robustness check against prior choice. Across the range of tested Beta priors, the posterior summaries preserve the same qualitative ordering, with the bMS consistently favoring a higher Ap-like fraction than the rMS, while all inferred fractions remain low in absolute terms. We therefore conclude that a small minority of eMS stars may show Ap-like spectrophotometric signatures, but Ap-like stars cannot account for the dominant population underlying the eMS phenomenon.

\subsection{Cluster-to-cluster consistency check}

To assess whether the combined inference is driven disproportionately by only one or two clusters, we repeat the same diagnostics for the full sample and for each of the eight clusters individually. Table~\ref{tab:cluster_check} summarizes both the star-by-star BF statistics and the inferred Ap-like fractions from the mixture modeling for the All, bMS, and rMS samples.

\begin{deluxetable*}{lcccccccc}
\tablecaption{Cluster-by-cluster consistency check from BF summaries and mixture-model Ap-like fractions\label{tab:cluster_check}}
\tablehead{
\colhead{Sample} &
\colhead{$N_{\rm all}$} &
\colhead{$f(BF>1)$} &
\colhead{$f(BF>3)$} &
\colhead{$f(BF>10)$} &
\colhead{median $\log BF$} &
\colhead{$\pi_{\rm All}$} &
\colhead{$\pi_{\rm bMS}$} &
\colhead{$\pi_{\rm rMS}$}
}
\startdata
Combined     & { 828} & { 22.6}\% & { 7.0}\%  & 1.9\% & { -0.49} & { 0.036}\,[0.016, { 0.066}] & { 0.079}\,[0.029, { 0.141}] & { 0.019}\,[{ 0.011}, { 0.049}] \\
NGC 2287   & 94  & 16.0\% & 2.1\%  & 1.1\% & -0.54 & 0.001\,[0.001, 0.056] & 0.001\,[0.001, 0.124] & 0.001\,[0.001, 0.086] \\
NGC 2516   & 103 & 16.5\% & 10.7\% & 2.9\% & -0.64 & 0.074\,[0.016, 0.169] & 0.189\,[0.044, 0.392] & 0.029\,[0.001, 0.119] \\
NGC 3532   & 220 & 30.0\% & 4.1\%  & 0.5\% & -0.29 & 0.001\,[0.001, 0.079] & 0.001\,[0.001, 0.109] & 0.019\,[0.001, 0.136] \\
M45        & 30  & 10.0\% & 0.0\%  & 0.0\% & -0.61 & 0.001\,[0.001, 0.106] & 0.001\,[0.001, { 0.407}] & 0.001\,[0.001, { 0.134}] \\
NGC 1039   & 33  & 33.3\% & 12.1\% & 3.0\% & -0.24 & 0.182\,[0.019, 0.475] & 0.287\,[0.001, 0.500] & 0.139\,[0.009, 0.492] \\
NGC 6475   & 79  & 13.9\% & 7.6\%  & 3.8\% & -0.92 & 0.036\,[0.001, 0.124] & 0.089\,[0.001, 0.289] & 0.001\,[0.001, 0.104] \\
Melotte 20 & 49  & 20.4\% & 6.1\%  & 0.0\% & -0.66 & 0.001\,[0.001, 0.101] & { 0.001},[0.001, { 0.181}] & 0.001\,[0.001, { 0.184}] \\
NGC 3114   & 236 & { 24.5}\% & { 10.5}\% & { 3.2}\% & { -0.54} & { 0.071}\,[{ 0.014, 0.149}] & 0.131\,[0.031, 0.267] & { 0.019}\,[0.001, { 0.119}] \\
\enddata
\tablenotetext{}{BF summary statistics are computed for the full sample of each cluster. The quantities $\pi_{\rm All}$, $\pi_{\rm bMS}$, and $\pi_{\rm rMS}$ are the maximum-likelihood estimates of the Ap-like fraction from the two-component mixture model, with 95\% confidence intervals shown in brackets. For the combined sample, the corresponding Bayesian posterior median ranges across the five adopted priors are 0.037--0.056 (All), 0.059--0.087 (bMS), and 0.020--0.051 (rMS), consistent with the same qualitative trend.}
\end{deluxetable*}

The cluster-by-cluster results show that most individual clusters provide only limited constraining power on their own. In many cases, the per-cluster subsamples are modest in size, so both the BF summaries and the inferred $\pi$ values carry substantial statistical uncertainty. This is particularly true for the rMS, whose inferred Ap-like fractions are generally low and often remain consistent with values close to zero within the uncertainties.

At the same time, the comparison across clusters reveals a qualitative pattern that broadly follows the combined result. In several clusters, the bMS tends to favor a somewhat larger Ap-like fraction than the rMS, whereas strong evidence for a substantial Ap-like contribution is generally lacking in either sequence. The contrast is clearest in NGC~2516 and NGC~3114, while NGC~1039 shows a similar tendency but with much weaker statistical support because of its small sample size. By contrast, clusters such as NGC~2287, NGC~3532, and M45 show only weak or near-zero signals, while Melotte~20 and NGC~6475 exhibit at most mild and statistically uncertain bMS enhancement.

Overall, these cluster-level results suggest that the combined inference is not obviously dominated by any single cluster. Although the uncertainties become substantially larger when the sample is split cluster by cluster, the inferred Ap-like fractions for most individual clusters remain low and are broadly consistent with the result. { With the exception of NGC~1039, the overall fractions are generally still below the \(\sim 40\%\)--\(50\%\) range often invoked in the literature for producing a distinct blue sequence \citep[according to \cite{Wang2022}, all bMS stars should be Ap-like. Here 40\%--50\% represents the number ratio of bMS stars, see][]{Li2017,Sun2019}. NGC~1039 only marginally overlaps this range at the edge of its confidence interval, while its central estimate remains much lower.} We therefore arrive at the same qualitative conclusion as from the combined analysis: Ap stars may contribute at a low level, and perhaps somewhat more strongly to the bMS than to the rMS, but they are unlikely to be the primary driver of the eMS phenomenon in these clusters. Future direct measurements of stellar magnetic fields for individual stars will be essential for overcoming the present statistical limitations and testing this interpretation more decisively.

\section{Discussion}

Our \(R_{5200}\)-based population analysis does not support a scenario in which Ap-like magnetic stars constitute the dominant slow-rotator population associated with the eMS/split-MS phenomenon. This result bears directly on the scenario proposed by \citet{Wang2022}, building on the merger-induced magnetism picture explored by \citet{Schneider2019}. In that framework, binary mergers can produce stars with strong, stable magnetic fields, which are then spun down by magnetic braking and may populate the bMS. A natural observational consequence is that the relevant population should show an enhanced incidence of Ap-like chemical peculiarity, for which the \(5200\,\mathrm{\AA}\) flux depression is a classical diagnostic \citep{Maitzen1976,Paunzen2005,Xie2025}. Across the eight clusters examined here, however, the color-defined eMS subsamples generally exhibit \(R_{5200}\) distributions that are much more similar to the normal-A reference population than to the Ap reference population, with at most a modest extension toward the Ap-like regime. Consistently, the mixture-model inference yields only a limited Ap-like component even for the bMS: in the combined sample, we find \(\pi_{\rm bMS}=0.076\) (95\% CI: \(0.029\)--\(0.139\)), compared with \(\pi_{\rm rMS}=0.021\) (95\% CI: \(0.004\)--\(0.051\)) and \(\pi_{\rm All}=0.039\) (95\% CI: \(0.016\)--\(0.069\)). Thus, while the bMS may host a somewhat elevated incidence of Ap-like stars relative to the rMS, the inferred fraction remains far too small to support a picture in which the eMS/split-MS slow-rotator population is dominated by stars with classical Ap-like chemical peculiarity. { This does not, however, exclude the \citet{Wang2022} scenario for individual stars. Some bMS stars may indeed have been spun down by strong fossil magnetic fields. Our constraint is instead demographic: stars with classical Ap-like chemical peculiarity are too rare to account for the bulk of the slow-rotator population associated with the eMS/split-MS phenomenon.}

This conclusion is genuinely constraining because the inferred Ap-like fractions remain low not only in the pooled analysis, but also in the cluster-by-cluster comparison. Previous split-MS interpretations generally require a slow-rotator fraction of at least \(15\%\)--\(25\%\), and often substantially more depending on the cluster and on how the bMS population is defined \citep[e.g.,][]{Milone2016,Milone2017,Li2017,Sun2019}. By contrast, our estimate for the full sample is only \(\pi_{\rm All}=0.039\) (95\% CI: \(0.016\)--\(0.069\)), well below even the low end of that range. The same overall picture persists when the clusters are considered individually: although the uncertainties become much larger because of the limited sample sizes, the best-fit \(\pi_{\rm All}\) values in most clusters still remain below \(\sim 15\%\). NGC 1039 is the main case that approaches this boundary, but its sample is small and the corresponding constraint is therefore weak. Taken together, these results indicate that the low Ap-like fraction is not an artifact driven by one or two particular clusters, but instead reflects a broadly consistent pattern across the present sample.

What our results constrain is not the general role of magnetism in the eMS/split-MS phenomenon, but specifically the prevalence of stars with classical Ap-like chemical peculiarity as traced by \(R_{5200}\). In this sense, the present analysis tests a concrete observational prediction of the \citet{Wang2022} scenario: if merger-induced magnetism and subsequent magnetic braking contribute substantially to the observed eMS morphology, then the corresponding population should show an enhanced incidence of Ap-like signatures. Our results do not support that interpretation. This distinction is important. A low inferred Ap-like fraction does not imply that magnetic effects are absent. Magnetism may still be present in only a minority of stars, or may manifest in forms that do not produce a strong Ap-like \(R_{5200}\) signature. Our results therefore argue against identifying the dominant slow-rotator population with classical Ap-like stars, but they should not be interpreted as ruling out magnetism more broadly as a contributing ingredient in the origin of the split main sequence.

Finally, several limitations of the present analysis should be kept in mind. Our use of \(R_{5200}\), Bayes factors, and mixture modeling is designed to constrain the Ap-like fraction at the population level, rather than to provide definitive classifications for individual stars. This is particularly important at the cluster level, where the sample sizes are often modest and the resulting uncertainties remain substantial. The current results therefore provide a statistical argument against a dominant contribution from classical Ap-like stars, but they do not by themselves establish the magnetic properties of any given star on the split main sequence. A more direct test will require spectropolarimetric measurements or other magnetic diagnostics for individual stars across both sequences. Such observations will be essential for determining whether magnetism plays only a secondary role, is confined to a minority population, or contributes in a form that is not traced by an Ap-like \(R_{5200}\) signature.

\section{Conclusions}
We have used \(R_{5200}\)-based population comparisons, Bayes-factor diagnostics, and mixture modeling to test whether the color-defined eMS populations in eight open clusters are dominated by stars with classical Ap-like chemical peculiarity, as would be expected under a natural extension of the \cite{Wang2022} scenario for the split main sequence. The \(R_{5200}\) distributions of the eMS subsamples are generally much more similar to those of the normal-A reference population than to those of the Ap reference population, and the inferred Ap-like fractions remain too small to support a picture in which classical Ap-like stars constitute the dominant slow-rotator population associated with the eMS/split-MS phenomenon.

This conclusion is supported both by the analysis and by the overall cluster-to-cluster comparison. At the same time, our results should be interpreted in the specific sense probed here: they constrain the prevalence of classical Ap-like chemical peculiarity as traced by \(R_{5200}\), rather than the broader role of magnetism itself. The present evidence therefore argues against identifying the dominant slow-rotator population with classical Ap-like stars, while leaving open the possibility that magnetism may still contribute in a more limited or less directly traceable way. More direct magnetic diagnostics, especially spectropolarimetric observations, will be needed to establish that role more definitively.
 
\begin{acknowledgments}

{ This work presents results from the European Space Agency (ESA) space mission Gaia. Gaia data are being processed by the Gaia Data Processing and Analysis Consortium (DPAC). Funding for the DPAC is provided by national institutions, in particular the institutions participating in the Gaia MultiLateral Agreement (MLA). The Gaia mission website is \url{https://www.cosmos.esa.int/gaia}. The Gaia archive website is \url{https://archives.esac.esa.int/gaia}.}

This project is supported by the National Natural Science Foundation of China (NSFC grant Nos. 12233013)
\end{acknowledgments}




%
\facilities{Gaia, AAT (HERMES)}

\software{astropy \citep{2013A&A...558A..33A,2018AJ....156..123A,2022ApJ...935..167A}}







\end{document}